\documentclass{Interspeech2024}

\usepackage{multirow}

\interspeechcameraready

\title{PPPR: Portable Plug-in Prompt Refiner for Text to Audio Generation}

\small \name[affiliation={1}]{Shuchen}{Shi}
\small \name[affiliation={2,*}]{Ruibo}{Fu}
\small \name[affiliation={2}]{Zhengqi}{Wen}
\small \name[affiliation={4}]{Jianhua}{Tao}
\small \name[affiliation={2}]{Tao}{Wang}
\small \name[affiliation={5}]{Chunyu}{Qiang}
\small \name[affiliation={2,3}]{Yi}{Lu}
\small \name[affiliation={2,3}]{Xin}{Qi} 
\small \name[affiliation={2}]{Xuefei}{Liu}
\small \name[affiliation={6}]{Yukun}{Liu}
\small \name[affiliation={2}]{Yongwei}{Li}
\small \name[affiliation={2,3}]{Zhiyong}{Wang}
\small \name[affiliation={2,3}]{Xiaopeng}{Wang}

\address{
  \small $^1$School of Computer and Information Engineering, Institute for Artificial Intelligence, Shanghai Polytechnic University\\
  \small $^2$Institute of Automation, Chinese Academy of Sciences
  \small $^3$School of Artificial Intelligence, Chinese Academy of Sciences
  \small $^4$Department of Automation and Beijing National Research Center for Information Science and Technology, Tsinghua University
  \small $^5$Tianjin University  
  \small $^6$University of Chinese Academy of Sciences
  \small{\thanks{* corresponding author}}
  }
\email{20221513084@stu.sspu.edu.cn, ruibo.fu@nlpr.ia.ac.cn}

\keywords{audio generation, large language model, chain-of-thought, diffusion model}

\begin{document}

\maketitle

\begin{abstract}
    
    Text-to-Audio (TTA) aims to generate audio that corresponds to the given text description, playing a crucial role in media production. The text descriptions in TTA datasets lack rich variations and diversity, resulting in a drop in TTA model performance when faced with complex text. To address this issue, we propose a method called Portable Plug-in Prompt Refiner, which utilizes rich knowledge about textual descriptions inherent in large language models to effectively enhance the robustness of TTA acoustic models without altering the acoustic training set. Furthermore, a Chain-of-Thought that mimics human verification is introduced to enhance the accuracy of audio descriptions, thereby improving the accuracy of generated content in practical applications. The experiments show that our method achieves a state-of-the-art Inception Score (IS) of 8.72, surpassing AudioGen, AudioLDM and Tango.
\end{abstract}

\section{Introduction}

With the development of Artificial Intelligence Generated Content (AIGC), related technologies such as Large Language Model (LLM) \cite{llm2023survey} and Latent Diffusion Model (LDM) \cite{diffusion2024survey,rombach2022ldm} have rapidly advanced. Text-to-speech (TTS) methods like VALL-E \cite{valle} and NaturalSpeech2 \cite{naturalspeech2} can synthesize natural and expressive speech. However, only clean speech \cite{fu2020tts1} cannot meet the current needs of AIGC creation. For instance, in virtual reality games, speech with scene sound effects is expected to be created to provide players with a more immersive experience. The text-to-audio (TTA) tasks can generate realistic scene sound effects based on text descriptions, playing an important role in generating speech that better fits the scene.

The TTA research revolves around two aspects, including acoustic models and textual descriptions. There are two approaches to building acoustic models in TTA tasks. One approach \cite{yang2023diffsound, kreuk2022audiogen} is to predict the discrete acoustic tokens of the target audio and then decode these tokens into an audio signal through a decoder. Another approach \cite{huang2023makeanaudio, liu2023audioldm, ghosal2023tango} is based on the LDM to predict continuous latent representations of audio. On the basis of good acoustic models, textual descriptions also play a crucial role in the final effect. The TTA tasks have evolved from generating audio based on single label \cite{labeltoaudio_1, labeltoaudio_2} to generating audio based on natural text descriptions \cite{yang2023diffsound, huang2023makeanaudio, kreuk2022audiogen, liu2023audioldm, ghosal2023tango}. In the early stages, TTA tasks primarily focused on generating sound based on single sound event label such as DogBark or Footstep. Compared to a single label, a natural text description is more effective in expressing people's needs. Natural text descriptions can provide a fine-grained description of the audio, including details such as the number of events, time sequence, sound scenes, and more \cite{liu2023audioldm}. With the high audio quality generated, the bottleneck constraining TTA at the application level lies primarily in the text descriptions. Therefore, deeper research into text descriptions in TTA is needed.

The insufficient diversity in text descriptions, as well as deficiencies in structured and regularized methods, hinder the development of TTA tasks. Previous studies have proposed some methods to alleviate this issue. AudioGen \cite{kreuk2022audiogen} and Tango \cite{ghosal2023tango} adopt a strategy based on mixed concatenation to increase the diversity of text descriptions. Meanwhile, they also mix the audio clips corresponding to the text descriptions in different ways. The mixed method increases the length of text descriptions and the complexity of audios, which is not conducive to the audio generation model learning the mapping between text descriptions and audios. Differing from the mixed method, Re-AudioLDM \cite{reaudioldm} proposes a retrieval-based method to enhance the diversity of text descriptions. The retrieval-based method has achieved good results, but it introduces additional conditional information to guide model generation, resulting in extra spend. The above methods do not consider increasing the diversity of text descriptions corresponding to the same audio clip, i.e. the fine-grained diversity of text descriptions. This may prevent the model from fully learning language expressions in different contexts. Models trained solely on limited text-audio pairs struggle to meet the descriptive forms found in real-world scenarios, leading to poor performance in practical application.

In this work, we propose Portable Plug-in Prompt Refiner (PPPR) TTA frontend augmentation method. PPPR enhances the robustness of the acoustic model by increasing the fine-grained diversity of text descriptions in the training dataset and improves the accuracy of the TTA model during practical applications by enhancing the accuracy of input text descriptions. Specifically, to address the issue of insufficient diversity in text descriptions, PPPR actively augments the original text descriptions using Llama \cite{touvron2023llama}, generating extensive text descriptions with fine-grained diversity. To improve the accuracy of text descriptions, PPPR uses the Chain of Thought (CoT) \cite{wei2022CoT} to guide Llama to gradually regularize input text descriptions, such as correcting spelling errors and checking the accuracy of descriptions, to obtain accurate regularized text descriptions. In summary, the contributions of this paper are as follows:

\begin{itemize}
    \item We leverage the knowledge of LLM to effectively address the narrow data issue in TTA tasks, achieving a complementary fusion of advantages between large and small models.
\begin{figure*}[t]
    \centering
    \includegraphics[width=1.8\columnwidth]{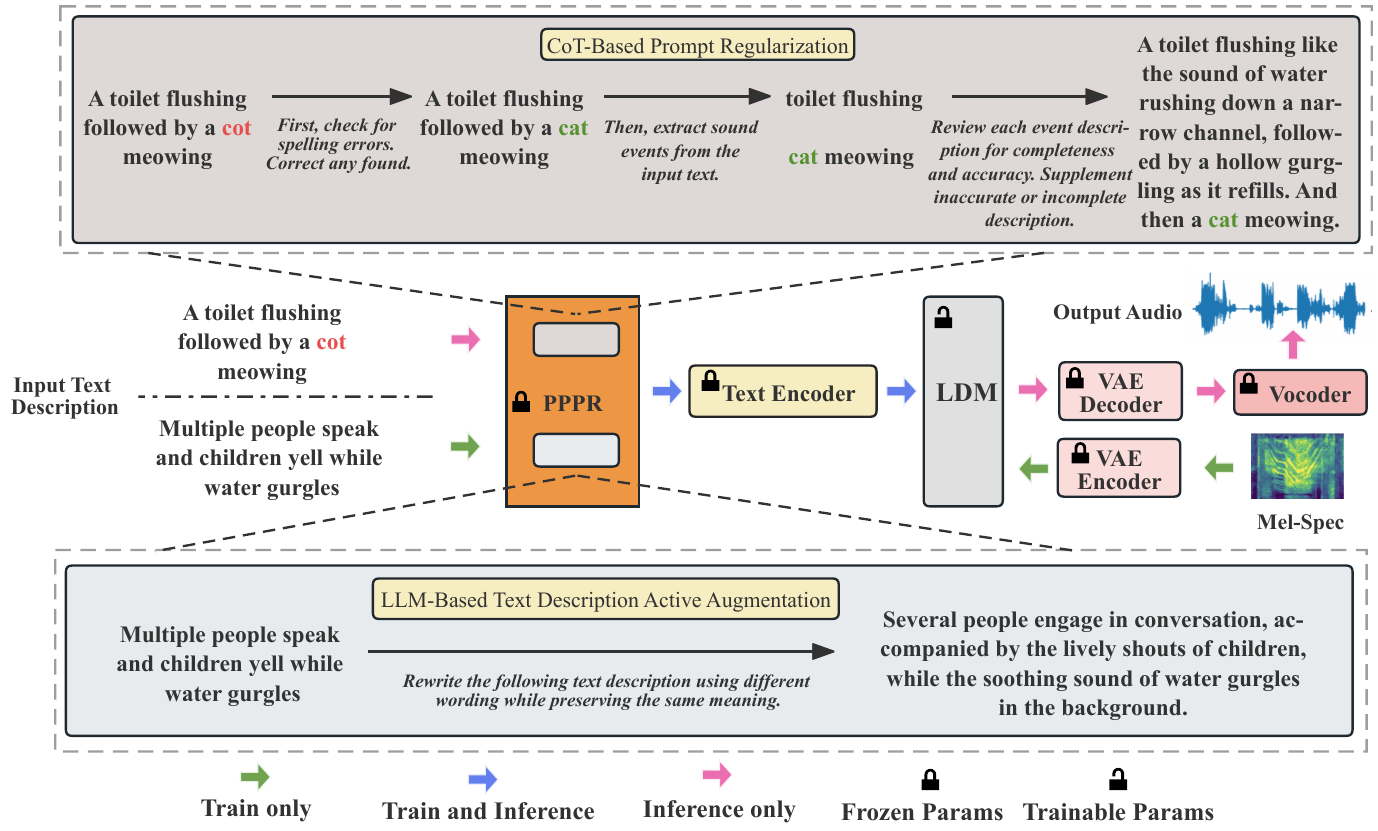}
    \caption{The process of training and inference for the audio generation model based on LDM using PPPR. The red words indicate words with spelling errors, while the green words indicate corrected words. The italicized words under the black unidirectional arrow indicate the prompts guiding Llama in processing at each step.}
    \label{fig:model_arc}
\end{figure*}
    
    \item We have increased the diversity of text descriptions in the TTA dataset and regularized the input text descriptions for the TTA model during the application phase.
    \item The model trained using PPPR's data diversity augmentation method performs better than baseline models trained using other augmentation methods. Regularization method in PPPR has improved the performance of our model and Tango.
\end{itemize}


\section{Method}
Illustrated in Figure \ref{fig:model_arc}, we enhance an audio generation model built on the LDM using the proposed method PPPR. The PPPR has two modules: 1) LLM-Based Text Description Active Augmentation; 2) CoT-Based Prompt Regularization. The audio generation model has four components: Text Encoder, LDM, Variational Autoencoder (VAE) \cite{kingma2013vae}, HiFi-GAN Vocoder \cite{kong2020hifigan}.
\subsection{LLM-Based Text Description Active Augmentation}
\label{sec:llm_aug}
Each audio clip can have multiple different text descriptions, representing the fine-grained diversity of text descriptions for the audio. By increasing this diversity, the exposure of audio clip to different text descriptions during training can be enhanced. Through increased exposure, the model can better learn the mapping between text descriptions and audio events. 

The annotation and comprehension capabilities of LLM have reached a level comparable to that of humans. PPPR actively increases the diversity of audio descriptions at a fine-grained level by leveraging LLM. We aim to increase the diversity of text descriptions by performing multiple rewriting iterations. Specifically, to generate rewritten text-audio pairs, it is necessary to first randomly select text-audio pairs from the AudioCaps \cite{kim2019audiocaps} dataset. Then, the designed prompt is inputted into LLM Llama. Llama rewrites the text from the randomly selected text-audio pairs based on the prompt.  It is crucial to ensure that the semantics of the text descriptions remain unchanged before and after rewriting; otherwise, rewriting will introduce noise into the training set. We have designed a prompt to assist Llama in rewriting while preserving the semantic meaning of the text descriptions. The specific prompt is as follows:
\begin{center}
\textit{Rewrite the following text description using different \\wording while preserving the same meaning.}
\end{center}

As shown in the Figure \ref{fig:model_arc}, the phrase "Multiple people speak" is rewritten as "Several people engage in conversation", carrying the same meaning. Therefore, the rewritten sentence retains the granularity of information from the original description. By repeatedly executing the above steps, we can increase the diversity of text descriptions.

\subsection{CoT-Based Prompt Regularization}
A TTA system should be able to accurately generate target audio based on input text descriptions. However, inaccuracies or incomplete descriptions in the text can lead to a decrease in accuracy of the generated audio. Trained on massive text data, LLM possesses excellent text processing capabilities under prompt guidance. The CoT enhances LLM's reasoning abilities, enabling it to better handle complex tasks.

PPPR utilizes CoT to gradually guide Llama in reasoning. Llama progressively adjusts the input text descriptions into more accurate and error-free expressions, thus enhancing the accuracy of the generation model. Specifically, by prompting "Reasoning with the following prompts step by step.", Llama is instructed to reason step by step. Firstly, a spelling check is required for the input text. Incorrect spelling may lead to semantic errors, causing the generation model to misunderstand the semantics of the text. Llama is fully capable of identifying and correcting errors. Next, extract the existing events from the text description and review each event description for completeness and accuracy. If any description is incomplete or inaccurate, supplement it to make the content more easily understood by the audio generation model. To implement these functionalities, we design the following prompt:
\begin{center}
\begin{flushleft}
\textit{
\quad Reasoning with the following prompts step by step.\\ 
\quad \quad1.First, check for spelling errors. Correct any found.\\
\quad \quad 2.Then, extract sound events from the input text. \\
\quad \quad 3.Review each event description for completeness and  \\ \quad \quad accuracy. Supplement inaccurate or incomplete\\ \quad \quad description.}
\end{flushleft}
\end{center}

As shown in the Figure \ref{fig:model_arc}, PPPR corrects spelling errors in the input using CoT, replacing the misspelled "cot" with "cat". Additionally, PPPR supplements the description that is not easily understandable. For instance, "A toilet flushing" is supplemented with "A toilet flushing like the sound of water rushing down a narrow channel, followed by a hollow gurgling as it refills". This enables the generation model to generate based on the supplemented description, even if it has not seen "the sound of toilet flushing" during training. Through these steps, the model's generation performance has been greatly improved.

\subsection{LLM-Baesd Refined Prompt Domain Text Encoder}

After being enhanced by LLM Llama, the diversity of text descriptions in the original training set has increased. The distribution of the enhanced text descriptions has become more complex, so using small pre-trained language models like BERT \cite{bert} is no longer sufficient to meet the demands. The FLAN-T5 model \cite{chung2022flan-t5} is pre-trained on a large-scale chain-of-thought-based and instruction-based dataset, making it suitable for complex text processing tasks. Therefore, we use the pre-trained LLM FLAN-T5-LARGE \cite{chung2022flan-t5} as the text encoder to obtain text embedding. Specifically, given input text T, Flan-T5-LARGE extracts its feature embedding $E^T$. $E^T$ serves as conditional information to guide the training of the generation model.

\subsection{LDM-Based Audio Generation Model}

\begin{table*}[ht!]
\centering
\small
\caption{The results on the Original AudioCaps test dataset with our method and the baseline models' method. The baseline models use mixed  concatenation methods, while we use the LLM-Based Text Descriptions Active Augmentation method in PPPR. And Ours does not use any augmentation methods. Audiogen uses the checkpoint of AudioGen-Medium-1.5B, which is trained on multiple datasets such as AudioCaps and AudioSet \cite{gemmeke2017audioset}, Clotho \cite{drossos2020clotho} and so on.}
\resizebox{0.8\textwidth}{!}{
\begin{tabular}{ccc|ccc|cc}
\toprule
\multirow{2}{*}{\textbf{Model}} & \multirow{2}{*}{\textbf{Datasets}} & \multirow{2}{*}{\textbf{Augmentation Methods}} & \multicolumn{3}{c|}{\textbf{Objective Metrics}} & \multicolumn{2}{c}{\textbf{Subjective Metrics}} \\ 
& & & FD~$\downarrow$ & KL~$\downarrow$ & IS~$\uparrow$ & OVL~$\uparrow$ & REL~$\uparrow$ \\
\midrule
Ground truth &$-$ & $-$ & $-$ & $-$ & $-$ & $85.56$ & $82.17$ \\
\midrule
AudioGen~\cite{kreuk2022audiogen} & $Multiple \hspace{4pt}datasets$ & $Mixed\hspace{5pt}Concatenation$  & $\mathbf{23.86}$ & $1.59$ & $8.59$ & $73.25$ & $70.76$ \\
AudioLDM~\cite{liu2023audioldm}  & $AudioCaps$ & $Mixed \hspace{5pt}Concatenation$  & $27.79$ & $1.65$ & $7.83$ & $70.15$ & $68.23$ \\

Tango \cite{ghosal2023tango} & $AudioCaps$  & $Mixed \hspace{5pt}Concatenation$   & $26.08$  & $\mathbf{1.38}$ & $8.04$ & $76.33$ & $74.22$ \\
Ours & $AudioCaps$ & $-$   & $29.40$  & $1.78$ & $6.33$ & $63.18$ & $60.69$ \\
\midrule
Ours-PPPR & $AudioCaps$ & $LLM-Based \hspace{5pt}Augmentation$  & $24.59$   & $1.39$ & $\mathbf{8.72}$ & $\mathbf{79.51}$ & $\mathbf{75.86}$ \\
\bottomrule
\end{tabular}
}
\label{tab:AudioCapsResults}
\end{table*}

The LDM \cite{rombach2022ldm} is used to generate intermediate latent features, with the feature embedding $\boldsymbol{E^T}$ extracted by Flan-T5-Large from the textual descriptions as conditions. During training, the LDM involves two processes: 1) A forward process in which the latent variable $z_0$, obtained by compressing mel-spectrogram $m$ using the VAE Encoder, gradually transforms into a standard Gaussian distribution $z_N$ over $N$ steps, with noise $\epsilon$ added at each step. 2) A reverse process for the model to predict the transition probabilities $\boldsymbol{\epsilon_\theta}$ of each step $\boldsymbol{n}$, for reconstruction the data $\boldsymbol{z_0}$ by removing the noise $\boldsymbol{z_N}$. The loss function \cite{rombach2022ldm} as: 
\begin{equation}
    L_{n}(\theta)={E}_{\boldsymbol{z}_{0},\boldsymbol{\epsilon},n}|| \boldsymbol{\epsilon} - \boldsymbol{\epsilon}_{\theta}(\boldsymbol{z}_{n},n,\boldsymbol{E^T})||^2_{2}
\label{ldm_loss}
\end{equation}
where $\boldsymbol{\epsilon_\theta}$ is the Gaussian distribution predicted by LDM with current state $\boldsymbol{z_n}$, current step $\boldsymbol{n}$, and current condition $\boldsymbol{E_T}$. 

We use VAE to compress spectrograms into latent features or to reconstruct spectrograms from latent features, and we use HiFi-GAN to reconstruct spectrograms into audio signals.

\section{Experiments}

\subsection{Datasets}
This work conducts experiments using the AudioCaps dataset, which comprises 38,679 audio clips in the training set, each paired with manually annotated captions. The validation set consists of 2,240 instances. Utilizing Llama, the audio caption corresponding to an audio clip in the training set is rewritten into four different sentences, resulting in five semantically equivalent but syntactically different captions per audio clip after rewriting. 

During training, one caption is randomly selected from the five captions corresponding to each audio clip to form a (text, audio) pair for training the LDM. Therefore, during training, the training set comprises a total of 193,395 unique instances, including 38,679 original manual annotations and 38,679 * 4 (PPPR Augmentaion) instances. For testing, the original  and the PPPR optimized AudioCaps test set will be used to evaluate the performance of the model respectively.

\subsection{Training Setting}
All datasets are resampled to 16kHz sampling rate and mono format, with samples padded to 10.24 seconds. We then extract mel-spectrograms from audios using parameters of 64 mel filter bands, 1024 window length, 1024 FFT, and 160 hop size, resulting in (1,64,1024) mel-spectrograms, akin to grayscale images with 64 height and 1024 width in 1 channels.
The FLAN-T5-LARGE text encoder is frozen in our setting and we only train the parameters of the latent diffusion model. The diffusion model is based on U-Net \cite{ronneberger2015unet} architecture and has a total of 866M parameters. We use 8 channels and a cross-attention \cite{vaswani2017attention} dimension of 1024 in the U-Net model.
We use the Adafactor optimizer \cite{shazeer2018adafactor} and AdafactorSchedule for training. We train the model for 16 epochs on the training dataset and report results for the checkpoint with the best validation loss. We use four 4090 GPUs for training, where it takes a total of 62 hours to train 16 epochs, with validation at the end of every epoch. We use a per GPU batch size of 4 with 4 gradient accumulation steps. The effective batch size for training is 4 (instance) * 4 (accumulation) * 4 (GPU)= 64.

\subsection{Baseline Models}
We compare our method with several mainstream methods in TTA tasks that employ mixed concatenation strategies for data augmentation: AudioGen \cite{kreuk2022audiogen}, AudioLDM \cite{liu2023audioldm}, and Tango \cite{ghosal2023tango}. AudioGen and Tango utilize a strategy that mixes audio clips and concatenates their corresponding audio captions. AudioLDM, trained solely on audio clips, only mixes audio clips.

\subsection{Evaluation Metrics}
In this work, we perform both objective evaluation and human subjective evaluation. The main metrics are used for objective evaluation include frechet distance (FD), inception score (IS), and kullback–leibler (KL) divergence. Analogous to the Frechet Inception Distance (FID) used in image synthesis, the FD score in audio domain quantifies the global similarity between created audio samples and the target samples without the need of using paired reference audio samples. The IS score is effective in evaluating both sample quality and diversity. The KL score is calculated using paired samples and it measures the divergence between two probability distributions. All of these three metrics are built upon a state-of-the-art audio classifier PANNs \cite{kong2020panns}. A higher IS score indicates a larger variety in the generated audio, while lower KL and FD scores indicate better audio quality. 
Following previous subjective evaluation method \cite{liu2023audioldm, ghosal2023tango} in TTA field, we ask five human evaluators to assess two aspects of the generated audio, including overall audio quality (OVL) and relevance to the text caption (REL). We randomly select 20 audio samples from each of baseline and proposed method generated, and ask participants to rate them on a scale from 1 to 100.
\section{Results and Analysis}

\begin{table}
\small
\caption{Using the CoT-Based Prompt Regularization method in PPPR optimizes the AudioCaps test set. The performance of Ours-PPPR and Tango on the test set. ”w/” indicates text descriptions are processed by the CoT-Based Prompt Regularization method in PPPR, while ”w/o” is the opposite.}
\centering
\begin{tabular}{cc|cc}
\toprule
\textbf{Model} & \textbf{Regularization} & \textbf{OVL}$\uparrow$ & \textbf{REL}$\uparrow$ \\ 

\midrule
Ground truth  & $-$ & $85.56$ & $82.17$ \\
\midrule
Tango  & $w/o$ & $76.33$ & $74.22$ \\
Tango  & $w/$ & $78.12$ & $75.94$ \\
Ours-PPPR  & $w/o$ & $79.51$ & $75.86$ \\
Ours-PPPR  & $w/$ & $\mathbf{82.53}$ & $\mathbf{77.39}$ \\
\bottomrule
\end{tabular}
\label{tab:CoT_Optimize_TsetSet}
\end{table}

\subsection{PPPR Enhances the Robustness of the Acoustic Model}
Table \ref{tab:AudioCapsResults} displays the main evaluation results on the original AudioCaps test set. It is worth noting that without using the PPPR to enhance the dataset during training, i.e., training on the original AudioCaps dataset, the audio generation model based on LDM does not demonstrate any advantage. Model trained using the proposed PPPR method achieves relatively good results in both objective and subjective evaluations. It obtains the highest score of 8.72 on the IS metric, indicating high diversity in the generated audio among all methods. Additionally, the scores on the FD and KL metrics are 24.59 and 1.39, respectively, which are not significantly different from the scores of 23.86 for AudioGen and 1.38 for Tango. Considering the scores of FD, KL, and IS, our method can produce diverse and high-quality audio. The proposed method also demonstrates very promising results in subjective evaluation, with an overall audio quality score of 79.51 and a relevance score of 75.86. We speculate that increasing the diversity of text descriptions at a fine-grained level has improved the robustness of acoustic models, thereby enhancing the performance of generation.

\begin{table}
\caption{Objective evaluation results for audio generation in the presence of multiple events or a single event in the text prompt in the AudioCaps test set.}
\resizebox{0.45\textwidth}{!}{
\begin{tabular}{cc|cc|cc}
\toprule
\multirow{2}{*}{\textbf{Model}} & \multirow{2}{*}{\textbf{Regularization}} & \multicolumn{2}{c|}{\textbf{Multi Events}} & \multicolumn{2}{c}{\textbf{Single Events}} \\ 
& & OVL~$\uparrow$ & REF~$\uparrow$ & OVL~$\uparrow$ & REL~$\uparrow$ \\
\midrule
Ground truth  & $-$ & $75.43$ & $73.08$ & $79.15$ & $78.50$ \\
\midrule
Tango  & $w/o$& $64.78$ & $63.25$ & $65.37$ & $64.15$ \\
Tango  & $w/$& $65.34$ & $64.13$ & $67.34$ & $65.31$ \\
Ours-PPPR   & $w/o$ & $67.82$ & $66.29$ & $69.34$ & $66.17$ \\
Ours-PPPR   & $w/$ & $\mathbf{68.10}$ & $\mathbf{66.79}$ & $\mathbf{70.95}$ & $\mathbf{67.55}$ \\
\bottomrule
\end{tabular}
}
\label{tab:multienevt}
\end{table}

\subsection{PPPR Enhances the Accuracy of Model in Generating Audio}
Table \ref{tab:CoT_Optimize_TsetSet} presents the test results after using PPPR to process the text descriptions. Subjective evaluation scores indicate that the generation performance of all models improves when the test text descriptions are regularized. Our method achieved the highest scores in the OVL and REF metrics, with scores of 82.53 and 77.39, respectively. We believe that after the text descriptions are regularized by PPPR, their accuracy is improved, thereby enhancing the accuracy of model in generating audio.

\subsection{PPPR is Helpful for Handling Multi-Events}
Furthermore, we analyzed how the PPPR method based on CoT performs when the test text descriptions contain multiple events. Consider the following examples: “Leaves rustling followed by a small bell chiming as birds chirp in the background” contains three separate sequential events, whereas “A duck quacks continuously” contains only one. We segregate the AudioCaps test set using the presence of temporal identifiers –when, while, before, after, then, follow, during – into two subsets, one with multiple events and the other with single event. We show the subjective evaluation results for audio generation on these subsets in Table \ref{tab:multienevt}. Our method achieves the best OVL and REF scores for both multiple events and single event instances. We infer that PPPR utilizes CoT to extract and process events from the descriptions one by one, enabling the information within the descriptions to be effectively decomposed and optimized, thus enhancing the accuracy of the entire generation system.

\section{Conclusion}

In this work, we propose the Portable Plug-in Prompt Refiner (PPPR) TTA front-end enhancement method. Specifically, we use LLM to enhance the diversity of text descriptions in the training dataset at a fine-grained level, thereby improving the robustness of the acoustic model. Additionally, we utilize CoT to regularize the text descriptions, enhancing the accuracy of the TTA model's outputs in practical applications. By using PPPR, we improve OVL and REF by 8\% and 4\% respectively compared to the best baseline model Tango. In future work, we will explore tasks involving joint generation of TTA and TTS.

\section{Acknowledgements}
This work is supported by the National Natural Science Foundation of China (NSFC) (No. 62101553, No. 62306316, No.
U21B20210, No. 62201571).


\bibliographystyle{IEEEtran}
\bibliography{main}

\begin{thebibliography}{10}
\providecommand{\url}[1]{#1}
\csname url@samestyle\endcsname
\providecommand{\newblock}{\relax}
\providecommand{\bibinfo}[2]{#2}
\providecommand{\BIBentrySTDinterwordspacing}{\spaceskip=0pt\relax}
\providecommand{\BIBentryALTinterwordstretchfactor}{4}
\providecommand{\BIBentryALTinterwordspacing}{\spaceskip=\fontdimen2\font plus
\BIBentryALTinterwordstretchfactor\fontdimen3\font minus \fontdimen4\font\relax}
\providecommand{\BIBforeignlanguage}[2]{{%
\expandafter\ifx\csname l@#1\endcsname\relax
\typeout{** WARNING: IEEEtran.bst: No hyphenation pattern has been}%
\typeout{** loaded for the language `#1'. Using the pattern for}%
\typeout{** the default language instead.}%
\else
\language=\csname l@#1\endcsname
\fi
#2}}
\providecommand{\BIBdecl}{\relax}
\BIBdecl

\bibitem{llm2023survey}
W.~X. Zhao, K.~Zhou, J.~Li, T.~Tang, X.~Wang, Y.~Hou, Y.~Min, B.~Zhang, J.~Zhang, Z.~Dong \emph{et~al.}, ``A survey of large language models,'' \emph{arXiv preprint arXiv:2303.18223}, 2023.

\bibitem{diffusion2024survey}
H.~Cao, C.~Tan, Z.~Gao, Y.~Xu, G.~Chen, P.-A. Heng, and S.~Z. Li, ``A survey on generative diffusion models,'' \emph{IEEE Transactions on Knowledge and Data Engineering}, pp. 1--20, 2024.

\bibitem{rombach2022ldm}
R.~Rombach, A.~Blattmann, D.~Lorenz, P.~Esser, and B.~Ommer, ``High-resolution image synthesis with latent diffusion models,'' in \emph{Proceedings of the IEEE/CVF conference on computer vision and pattern recognition}, 2022, pp. 10\,684--10\,695.

\bibitem{valle}
C.~Wang, S.~Chen, Y.~Wu, Z.~Zhang, L.~Zhou, S.~Liu, Z.~Chen, Y.~Liu, H.~Wang, J.~Li, L.~He, S.~Zhao, and F.~Wei, ``Neural codec language models are zero-shot text to speech synthesizers,'' \emph{CoRR}, vol. abs/2301.02111, 2023.

\bibitem{naturalspeech2}
K.~Shen, Z.~Ju, X.~Tan, E.~Liu, Y.~Leng, L.~He, T.~Qin, sheng zhao, and J.~Bian, ``Naturalspeech 2: Latent diffusion models are natural and zero-shot speech and singing synthesizers,'' in \emph{The Twelfth International Conference on Learning Representations}, 2024.

\bibitem{fu2020tts1}
R.~Fu, J.~Tao, Z.~Wen, J.~Yi, and T.~Wang, ``Focusing on attention: prosody transfer and adaptative optimization strategy for multi-speaker end-to-end speech synthesis,'' in \emph{ICASSP 2020-2020 IEEE International Conference on Acoustics, Speech and Signal Processing (ICASSP)}.\hskip 1em plus 0.5em minus 0.4em\relax IEEE, 2020, pp. 6709--6713.

\bibitem{yang2023diffsound}
D.~Yang, J.~Yu, H.~Wang, W.~Wang, C.~Weng, Y.~Zou, and D.~Yu, ``Diffsound: Discrete diffusion model for text-to-sound generation,'' \emph{IEEE/ACM Transactions on Audio, Speech, and Language Processing}, 2023.

\bibitem{kreuk2022audiogen}
F.~Kreuk, G.~Synnaeve, A.~Polyak, U.~Singer, A.~D{\'e}fossez, J.~Copet, D.~Parikh, Y.~Taigman, and Y.~Adi, ``Audiogen: Textually guided audio generation,'' in \emph{The Eleventh International Conference on Learning Representations}, 2023.

\bibitem{huang2023makeanaudio}
R.~Huang, J.~Huang, D.~Yang, Y.~Ren, L.~Liu, M.~Li, Z.~Ye, J.~Liu, X.~Yin, and Z.~Zhao, ``Make-an-audio: Text-to-audio generation with prompt-enhanced diffusion models,'' in \emph{International Conference on Machine Learning}.\hskip 1em plus 0.5em minus 0.4em\relax PMLR, 2023, pp. 13\,916--13\,932.

\bibitem{liu2023audioldm}
H.~Liu, Z.~Chen, Y.~Yuan, X.~Mei, X.~Liu, D.~P. Mandic, W.~Wang, and M.~. Plumbley, ``Audioldm: Text-to-audio generation with latent diffusion models,'' in \emph{International Conference on Machine Learning}, 2023.

\bibitem{ghosal2023tango}
D.~Ghosal, N.~Majumder, A.~Mehrish, and S.~Poria, ``Text-to-audio generation using instruction guided latent diffusion model,'' in \emph{Proceedings of the 31st ACM International Conference on Multimedia}, 2023, pp. 3590--3598.

\bibitem{labeltoaudio_1}
X.~Liu, T.~Iqbal, J.~Zhao, Q.~Huang, M.~D. Plumbley, and W.~Wang, ``Conditional sound generation using neural discrete time-frequency representation learning,'' in \emph{2021 IEEE 31st International Workshop on Machine Learning for Signal Processing (MLSP)}.\hskip 1em plus 0.5em minus 0.4em\relax IEEE, 2021, pp. 1--6.

\bibitem{labeltoaudio_2}
Y.~Yuan, H.~Liu, X.~Liu, X.~Kang, P.~Wu, M.~D. Plumbley, and W.~Wang, ``Text-driven foley sound generation with latent diffusion model,'' \emph{arXiv preprint arXiv:2306.10359}, 2023.

\bibitem{reaudioldm}
Y.~Yuan, H.~Liu, X.~Liu, Q.~Huang, M.~. Plumbley, and W.~Wang, ``Retrieval-augmented text-to-audio generation,'' \emph{ArXiv}, vol. abs/2309.08051, 2023.

\bibitem{touvron2023llama}
H.~Touvron, T.~Lavril, G.~Izacard, X.~Martinet, M.-A. Lachaux, T.~Lacroix, B.~Rozi{\`e}re, N.~Goyal, E.~Hambro, F.~Azhar \emph{et~al.}, ``Llama: Open and efficient foundation language models,'' \emph{arXiv preprint arXiv:2302.13971}, 2023.

\bibitem{wei2022CoT}
J.~Wei, X.~Wang, D.~Schuurmans, M.~Bosma, F.~Xia, E.~Chi, Q.~V. Le, D.~Zhou \emph{et~al.}, ``Chain-of-thought prompting elicits reasoning in large language models,'' \emph{Advances in Neural Information Processing Systems}, vol.~35, pp. 24\,824--24\,837, 2022.

\bibitem{kingma2013vae}
D.~P. Kingma and M.~Welling, ``Auto-encoding variational bayes,'' \emph{CoRR}, vol. abs/1312.6114, 2013.

\bibitem{kong2020hifigan}
J.~Kong, J.~Kim, and J.~Bae, ``Hifi-gan: Generative adversarial networks for efficient and high fidelity speech synthesis,'' \emph{Advances in Neural Information Processing Systems}, vol.~33, pp. 17\,022--17\,033, 2020.

\bibitem{kim2019audiocaps}
C.~D. Kim, B.~Kim, H.~Lee, and G.~Kim, ``Audiocaps: Generating captions for audios in the wild,'' in \emph{Proceedings of the 2019 Conference of the North American Chapter of the Association for Computational Linguistics: Human Language Technologies, Volume 1 (Long and Short Papers)}, 2019, pp. 119--132.

\bibitem{bert}
J.~Devlin, M.-W. Chang, K.~Lee, and K.~Toutanova, ``Bert: Pre-training of deep bidirectional transformers for language understanding,'' \emph{arXiv preprint arXiv:1810.04805}, 2018.

\bibitem{chung2022flan-t5}
H.~W. Chung, L.~Hou, S.~Longpre, B.~Zoph, Y.~Tay, W.~Fedus, Y.~Li, X.~Wang, M.~Dehghani, S.~Brahma \emph{et~al.}, ``Scaling instruction-finetuned language models,'' \emph{arXiv preprint arXiv:2210.11416}, 2022.

\bibitem{gemmeke2017audioset}
J.~F. Gemmeke, D.~P. Ellis, D.~Freedman, A.~Jansen, W.~Lawrence, R.~C. Moore, M.~Plakal, and M.~Ritter, ``Audio set: An ontology and human-labeled dataset for audio events,'' in \emph{2017 IEEE international conference on acoustics, speech and signal processing (ICASSP)}.\hskip 1em plus 0.5em minus 0.4em\relax IEEE, 2017, pp. 776--780.

\bibitem{drossos2020clotho}
K.~Drossos, S.~Lipping, and T.~Virtanen, ``Clotho: An audio captioning dataset,'' in \emph{ICASSP 2020-2020 IEEE International Conference on Acoustics, Speech and Signal Processing (ICASSP)}.\hskip 1em plus 0.5em minus 0.4em\relax IEEE, 2020, pp. 736--740.

\bibitem{ronneberger2015unet}
O.~Ronneberger, P.~Fischer, and T.~Brox, ``U-net: Convolutional networks for biomedical image segmentation,'' in \emph{Medical Image Computing and Computer-Assisted Intervention--MICCAI 2015: 18th International Conference, Munich, Germany, October 5-9, 2015, Proceedings, Part III 18}.\hskip 1em plus 0.5em minus 0.4em\relax Springer, 2015, pp. 234--241.

\bibitem{vaswani2017attention}
A.~Vaswani, N.~Shazeer, N.~Parmar, J.~Uszkoreit, L.~Jones, A.~N. Gomez, {\L}.~Kaiser, and I.~Polosukhin, ``Attention is all you need,'' \emph{Advances in neural information processing systems}, vol.~30, 2017.

\bibitem{shazeer2018adafactor}
N.~Shazeer and M.~Stern, ``Adafactor: Adaptive learning rates with sublinear memory cost,'' in \emph{International Conference on Machine Learning}.\hskip 1em plus 0.5em minus 0.4em\relax PMLR, 2018, pp. 4596--4604.

\bibitem{kong2020panns}
Q.~Kong, Y.~Cao, T.~Iqbal, Y.~Wang, W.~Wang, and M.~D. Plumbley, ``Panns: Large-scale pretrained audio neural networks for audio pattern recognition,'' \emph{IEEE/ACM Transactions on Audio, Speech, and Language Processing}, vol.~28, pp. 2880--2894, 2020.

\end{thebibliography}

\end{document}